\documentclass[sigconf]{acmart}

\makeatletter

\usepackage{url}
\usepackage{xcolor}
\usepackage{braket}

\usepackage{booktabs}
\usepackage{graphicx}
\usepackage{subcaption}
\usepackage{algorithm}
\usepackage{algcompatible}

\usepackage{listings}
\usepackage{color}
\usepackage{courier}
\usepackage{hyperref}

\definecolor{mygreen}{rgb}{0,0.6,0}
\definecolor{mygray}{rgb}{0.5,0.5,0.5}
\definecolor{mymauve}{rgb}{0.58,0,0.82}
\newcommand{\qaoakit}{\texttt{QAOAKit}}
\newcommand{\qaoajl}{\texttt{QAOA.jl}}
\newcommand{\juliq}{\texttt{JuliQAOA}}

\copyrightyear{2023} 
\acmYear{2023} 
\setcopyright{rightsretained} 
\acmConference[SC-W 2023]{Workshops of The International Conference on High Performance Computing, Network, Storage, and Analysis}{November 12--17, 2023}{Denver, CO, USA}
\acmBooktitle{Workshops of The International Conference on High Performance Computing, Network, Storage, and Analysis (SC-W 2023), November 12--17, 2023, Denver, CO, USA}
\acmDOI{10.1145/3624062.3624220}
\acmISBN{979-8-4007-0785-8/23/11}

\begin{document}

\title{JuliQAOA: Fast, Flexible QAOA Simulation}

\author{John Golden}
\affiliation{%
  \institution{Los Alamos National Laboratory}
  \city{Los Alamos}
  \state{NM}
  \country{USA}
  \postcode{87545}
}

\author{Andreas Baertschi}
\affiliation{%
  \institution{Los Alamos National Laboratory}
  \city{Los Alamos}
  \state{NM}
  \country{USA}
  \postcode{87545}
}
\author{Daniel O'Malley}
\affiliation{%
  \institution{Los Alamos National Laboratory}
  \city{Los Alamos}
  \state{NM}
  \country{USA}
  \postcode{87545}
}
\author{Elijah Pelofske}
\affiliation{%
  \institution{Los Alamos National Laboratory}
  \city{Los Alamos}
  \state{NM}
  \country{USA}
  \postcode{87545}
}
\author{Stephan Eidenbenz}
\affiliation{%
  \institution{Los Alamos National Laboratory}
  \city{Los Alamos}
  \state{NM}
  \country{USA}
  \postcode{87545}
}

\begin{abstract}
We introduce \juliq{}, a simulation package specifically built for the Quantum Alternating Operator Ansatz (QAOA). \juliq{} does not require a circuit-level description of QAOA problems, or another package to simulate such circuits, instead relying on a more direct linear algebra implementation. This allows for increased QAOA-specific performance improvements, as well as improved flexibility and generality. \juliq{} is the first QAOA package designed to aid in the study of both constrained and unconstrained combinatorial optimization problems, and can easily include novel cost functions, mixer Hamiltonians, and other variations. \juliq{} also includes robust and extensible methods for learning optimal angles. Written in the Julia language, \juliq{} outperforms existing QAOA software packages and scales well to HPC-level resources. \juliq{} is available at \url{https://github.com/lanl/JuliQAOA.jl}. 
\end{abstract}


\maketitle

\section{Introduction}
\label{section:Introduction}
The Quantum Alternating Operator Ansatz~\cite{hadfield_qaoa} (QAOA), building on the earlier Quantum Approximate Optimization Algorithm~\cite{Farhi2014, farhi2015quantum}, is a leading quantum algorithm for solving combinatorial optimization problems.
Many questions remain open regarding the overall power of QAOA, as there exist few theoretical guarantees on performance. 
Instead, QAOA is most commonly studied as a heuristic optimization tool, involving both a classical outer loop and quantum inner loop.
The building blocks of QAOA are an optimization problem, encoded in a cost function $C(x)$ on binary strings ($x$), an initial state $\ket{\psi_0}$, a unitary determined by the mixer Hamiltonian $H_M$, and a set of $2p$ parameters $\{\beta_i, \gamma_i\}$ known as angles.
Finally, the cost function $C(x)$ is used to create a cost, or phase separator Hamiltonian, traditionally of the form $H_C\ket{x} = C(x)\ket{x}$.
These elements combine to form a $p$-round QAOA: using the state
\begin{equation}
	\ket{\boldsymbol{\beta}, \boldsymbol{\gamma}} =e^{-i\beta_p H_M}e^{-i \gamma_p H_C}\ldots e^{-i\beta_1 H_M}e^{-i \gamma_1 H_C}\ket{\psi_0},
\end{equation}
one uses classical optimization techniques to find $\{\beta_i, \gamma_i\}$ which maximize (or minimize) $\braket{\boldsymbol{\beta}, \boldsymbol{\gamma}|C(x)|\boldsymbol{\beta}, \boldsymbol{\gamma}}$.

The physical intuition behind QAOA is that the the mixer Hamiltonian is designed to generate destructive interference between states with poor $C(x)$ and constructive interference between states with good $C(x)$.
QAOA can also be viewed as a Trotterization of quantum annealing, with $H_C$ serving as the initial Hamiltonian and the $H_M$ as the target.

\begin{figure}[t]
  \centering
  \includegraphics[width=0.48\textwidth]{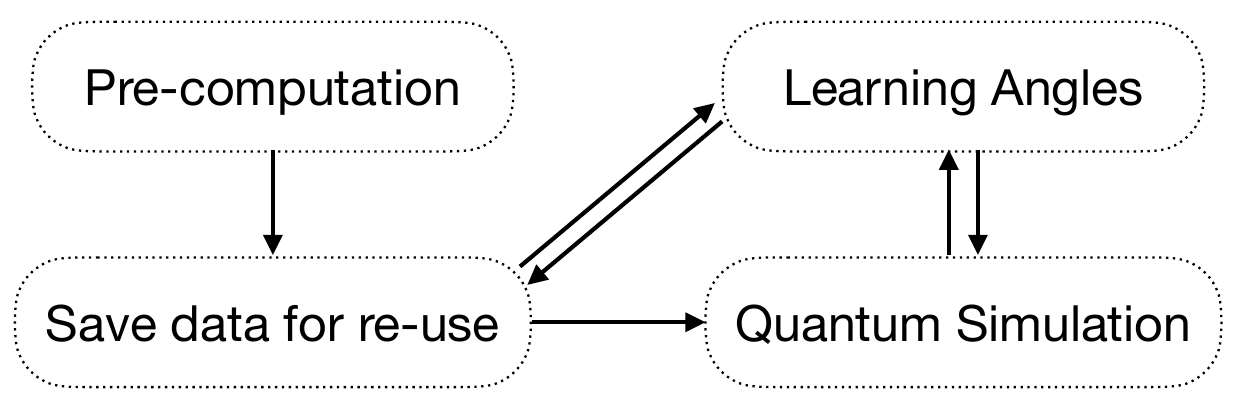}
  \caption{Conceptual overview of \juliq{}. To begin evaluating a QAOA, \juliq{} precomputes the value of the cost function $C(x)$ across all feasible states, as well as a diagonalized version of the mixer Hamiltonian. The quantum simulation subroutine calls on this information to efficiently compute $\langle C(x) \rangle$. The angle-finding outer-loop uses $\langle C(x) \rangle$ to find optimal angles, saving results for intermediate steps.}
  \label{fig1}
\end{figure}

While there are some theoretical proofs on the efficacy of QAOA~\cite{Farhi2014}, the ideal choices for many of the above quantities -- $\ket{\psi_0}$, $H_M$, classical optimization technique -- for a given $C(x)$ remain the focus of much active research.
Numerical experimentation is therefore a critical tool in evaluating the potential and limitations of QAOA.
Previous numerical methods for testing QAOA have largely been based off of general purpose quantum circuit simulators~\cite{Shaydulin_2021,Bode2023}, which are often costly to execute and have thus limited simulations to small numbers of qubits and few rounds (e.g. $n<10, p<5$).
Meanwhile, purpose-built QAOA simulation has pushed up to $n=30$, but only in the special case of MaxCut on 3-regular graphs~\cite{Lykov_2021}. 
Here we introduce the QAOA simulation tool \juliq{}, which has been specifically built to study a wide range of QAOA problems.
\juliq{} enables quick, low-overhead QAOA simulation on personal computers, while also scaling well to HPC-level resources. 
\juliq{} has already enabled robust numerical studies in several publications~\cite{golden2023quantum, pelofske2023highround, golden2022evidence, golden2021thresholdbased}, and will be released as an open source package later this year to facilitate further development.

\section{\juliq{} Overview}
\label{section:Overview}
\begin{figure}[t]
  \centering
  \includegraphics[width=0.48\textwidth]{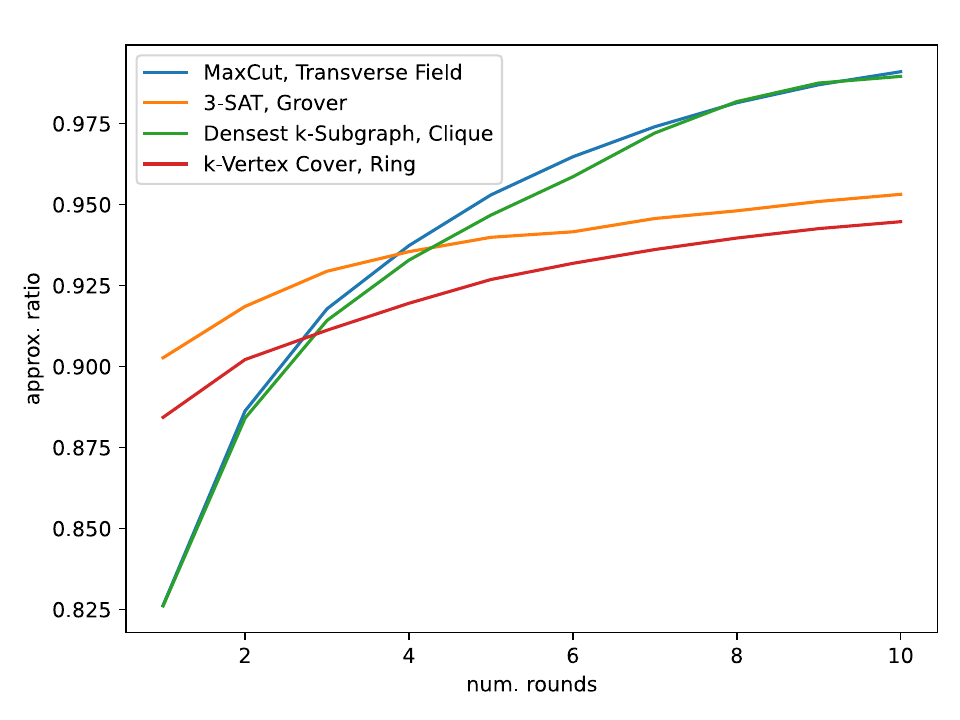}
  \caption{Example use of \juliq{} to find high-quality angles for increasing $p$ across a variety of problem types with different mixers. Each line represents the results for a single random instance of the given problem type. All problems have $n=12$. The MaxCut, Densest $k$-Subgraph, and $k$-Vertex Cover instances are all on a random $G(n,0.5)$ graph, with $k=6$ for the constrained problems. The 3-SAT instance has clause density 6. The mixers are described in Section \ref{section:pre_computation}, and the angle-finding technique is described in Section \ref{section:Angle_Finding}. }
  \label{fig:survey}
\end{figure}

\juliq{} is a quantum simulator expressly built for exact statevector simulation of QAOA, exploiting several QAOA-specific features in order to reduce time and memory usage.
In particular, the repeated nature of the QAOA algorithm means that several key components can be pre-calculated and re-used throughout the simulation.
\juliq{} is designed for large-scale and wide-ranging numerical experimentation, incorporating a broad variety of cost and mixer Hamiltonians. 
See Figure~\ref{fig1} for a conceptual overview.
The core distinguishing features of \juliq{} are:
\begin{itemize}
    \item \textbf{No circuits} -- does not require a circuit-level description of either the cost or mixer Hamiltonians.
    \item \textbf{Flexible} -- can efficiently study both constrained and unconstrained optimization problems, and incorporate non-traditional QAOA approaches (e.g. multi-angle~\cite{herrman2021multiangle}, threshold phase separator~\cite{golden2021thresholdbased}, different initial states~\cite{egger2021warm}).
    \item \textbf{Performant} -- outperforms other QAOA implementations in terms of time and memory.
    \item \textbf{HPC-friendly} -- easy to use on large scale HPC clusters with either GPUs or CPUs.
\end{itemize}
As an example of the capabilities of \juliq{}, Figure~\ref{fig:survey} shows the results of using \juliq{} to find high-quality angles for random instances of four different problem types (MaxCut, 3-SAT, Densest $k$-Subgraph, $k$-Vertex Cover), each with a different mixer (Transverse Field, Grover, Clique, Ring), at $n=12$ qubits up to $p=10$ rounds.
The data for Figure~\ref{fig:survey} was generated on an Apple M2 Max laptop in $\le 1$hr.

The flexibility and performance of \juliq{} comes from three main sources: pre-computation, efficient quantum simulation, and robust angle-finding.

\subsection{Pre-Computation}
\label{section:pre_computation}
The pre-computation step is used to store key quantities which will be re-used throughout the calculation.
First, for a given optimization problem characterized by a cost function $C(x)$, the user computes and stores the value of $C(x)$ across all feasible states $S$.
In the case of unconstrained optimization, $S$ is the set of all $n$-qubit computational basis states.
For constrained optimization $S$ is a subspace, e.g., in the cases of $k$-Densest Subgraph or Max $k$-Vertex Cover the feasible solutions are states with $k$ qubits set to 1 (also known as the Dicke state $\ket{D^n_k}$, which is an equal weight superposition of all $n$ qubit states with Hamming Weight $k$~\cite{PhysRev.93.99, B_rtschi_2019, 9951196}).
The feasible subspace thus has size $\binom{n}{k}$, and one need only provide $C(x)$ evaluated on that subspace.

The second quantity in the pre-computation step is the mixer Hamiltonian.
For unconstrained problems, \juliq{} is specifically optimized to use mixer Hamiltonians written as  sums of products Pauli $X$ operators.
This choice covers a broad array of mixers \cite{golden2023quantum}, including the original transverse-field mixer~\cite{Farhi2014, hadfield_qaoa} and Grover mixer~\cite{baertschi2020grover}.
Time evolution of such a mixer Hamiltonian, which we generically denote $f(X_i)$ can be diagonalized by exploiting $HZH = X$: 
\begin{equation}
    e^{-i \beta f(X_i)} = H^{\otimes n} e^{-i \beta f(Z_i)}H^{\otimes n}.
\end{equation}
This reduces the complicated matrix exponential into a simple sequence of single-qubit operations and vector exponentiation.
For a given input set of $X_i$ mixing terms, we pre-compute and store the equivalent diagonal matrix in terms of $Z_i$.

For constrained problems, we employ a similar trick of diagonalization. 
Commonly studied mixers for constrained problems, e.g. the Clique $\sum_{i<j} X_i X_j + Y_i Y_j$ and Ring $\sum_{i = j+1} X_i X_j + Y_i Y_j$ mixers~\cite{hadfield_qaoa}, are designed to only mix between feasible states, however they generally do not admit a diagonalization in terms of single-qubit gates.
Instead, we pre-compute the eigenvalue decomposition $H_M = VDV^{-1}$, where $D$ is a diagonal matrix.
These computations can be costly but need only be done once, and then the results are stored for future re-use.
This again allows us to avoid matrix exponentiation and instead use $e^{-i \beta H_M} = V e^{-i \beta D}V^{-1}$.
As with the cost function, we do not consider the mixer as a $2^n \times 2^n$ matrix, instead restricting the action of the mixer to the feasible subspace, e.g. resulting in matrices of size $\binom{n}{k} \times \binom{n}{k}$ for Dicke state problems.
Furthermore, any mixer that is not of the above formats (for both constrained and unconstrained problems) can be implemented as a unitary matrix, and \juliq{} will compute and store the eigendecomposition.

\subsection{Simulation}
\label{section:simulation}
Once the cost function and mixer have been pre-computed, simulating the QAOA is a relatively straightforward effort in efficient linear algebra.
\juliq{} is written in Julia \cite{Julia-2017}, a language that is inter-operable with Python at performance levels comparable to C.
GPU support is currently enabled via \texttt{CUDA.jl} \cite{besard2018juliagpu}, and was used in~\cite{golden2022evidence} on an NVIDIA RTX A6000 with 48GB to analyze constrained optimization problems with $n=18$ up to $p \approx 30$.
The main limiting factor in this case was the memory requirements in finding the eigendecomposition of the Clique mixer matrix.

In calculating the statevector simulation, we pre-allocate and re-use memory, allowing for functionally zero overhead.
As discussed previously, the most time-intensive part of the simulation is multiplying by the $V$ and $V^{-1}$ of the eigendecomposition, $H_M = VDV^{-1}$. 
In the case of the Clique and Ring mixers, this is necessarily a matrix-level operation (unless one adopts Trotter approximations, which we do not consider here).

In the case of unconstrained problems with Pauli $X$ mixers, $V = V^{-1} = H^{\otimes n}$. 
Because this is a sequence of single qubit operations, they can be efficiently implemented in $O(2^n)$ time via appropriate tensor contractions.
The specifics of our Julia implementation are drawn from \texttt{Yao.jl}~\cite{YaoFramework2019}, the same basic techniques have been used elsewhere~\cite{gitproj, Sack_2021}.

\subsection{Angle Finding}
\label{section:Angle_Finding}
Finally, we include an extensible format for learning good angles for QAOA in the classical angle-finding outer-loop.
\juliq{} has been specifically designed to exploit automatic differentiation~\cite{baydin2018automatic}, commonly referred to as autodiff or AD.
AD is a method of calculating exact gradients of complicated functions, in essence using the chain rule across every step of the function.
\juliq{} relies on \texttt{Enzyme.jl}~\cite{10.1145/3458817.3476165,NEURIPS2020_9332c513,10.5555/3571885.3571964}, which works with code compiled at the LLVM level.
This allows for highly efficient AD, and also avoids difficulties with complex numbers as well as in-place memory modification, common issues in other AD packages such as \texttt{ForwardDiff.jl} and \texttt{ReverseDiff.jl}.
We discuss the specifics of the performance gains garnered by AD in Section \ref{section:Performance}.

\begin{figure}[t]
  \centering
  \includegraphics[width=0.48\textwidth]{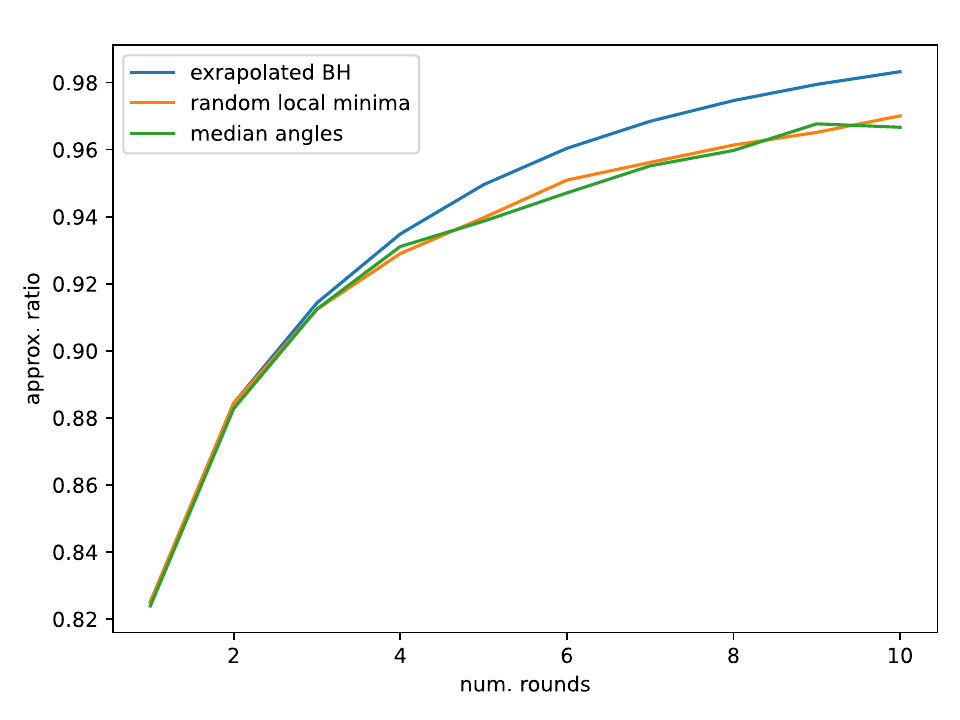}
  \caption{Comparison of our extrapolated basinhopping angle-finding technique against the random local minima exploration and median angles approaches of~\cite{Lotshaw_2021}. The speed-up provided by \juliq{} allows us to easily extend the analysis of~\cite{Lotshaw_2021}, which was limited to $n=9, p=3$. These results are mean values across 50 random MaxCut instances at $n=12$ on $G(n,0.5)$ graphs.}
  \label{fig:angle_finding}
\end{figure}

Research done with early versions of \juliq{}~\cite{golden2022evidence} showed the power of iterative angle-finding, i.e. using high-quality angles for a $(p-1)$-round QAOA to seed angles for the $p$-round QAOA.
Starting with these extrapolated angles, we then use the basinhopping algorithm~\cite{Wales_1997} to explore nearby local minima.
Other common angle-finding methods are grid search, random local minima exploration, and median angles.
For example, in~\cite{Lotshaw_2021} they use all three of these techniques to study MaxCut with the Transverse Field mixer up to $n=9$ and $p=3$.
The random local minima search begins at a random choice of angles and then uses the Broyden-Fletcher-Goldfarb-Shanno (BFGS)~\cite{Flet87} algorithm to find the closest local minima.
This process is repeated 100 times, each from a different random initial point, and the lowest minima is taken as the final output.
The median angles approach takes the results of the random local minima approach over a large number of problem instances and then finds the median angles.
In Figure~\ref{fig:angle_finding} we extend this analysis to $n=12$ and up to $p=10$, and show the difference in performance against our extrapolated basinhopping approach.

\juliq{} has a built-in wrapper for this iterative approach that saves angles for each round as they are found. User-defined methods for exploring angles are also supported.

\subsection{Grover Mixer}
\label{section:Grover_Mixer}
\juliq{} provides an additional level of specialized and optimized code when studying the Grover mixer~\cite{baertschi2020grover}.
The Grover mixer is given by $H_G = \ket{\psi_0}\bra{\psi_0}$, where $\ket{\psi_0}$ is the equal superposition over all computational basis states.
The Grover mixer has several interesting properties.
First, it conserves Hamming weight when mixing, making it suitable for Hamming weight constrained as well as unconstrained problems. 
Second, it can be used in conjunction with a threshold-based phase separator to reproduce Grover's search algorithm as a QAOA~\cite{golden2022evidence}.
Third, it gives fair sampling, that is, all states with equal objective values have the same amplitude.
The third property can be significantly exploited to calculate the expectation value in a recursive fashion~\cite{Golden_2021}.

\juliq{} includes special code for evaluating Grover-QAOA simulations, allowing simulation for very large (up to $n=100$) problems.
For problems of this size, significant computational bottlenecks arise in the traditionally straightforward step of pre-computing the $O(2^{100})$ objective values.
Such calculations are made possible by the fact that, in the special case of the Grover mixer, one does not need to store all objective values, but instead only the distinct values that the objective function can take as well as their degeneracies.
The calculation of these degeneracies can be easily spread across many threads or GPUs.
Determining which binary states are evaluated on each worker is simple in the case of unconstrained problems, as the range of integers $1\ldots2^n$ can be partitioned appropriately.
In the case of Hamming weight $k$ states, one can use Gosper's hack~\cite{gosper} to efficiently iterate through all binary strings with $k$ ones.

\section{Examples}

\paragraph{Note:} Please see the latest documentation, available at \url{https://lanl.github.io/JuliQAOA.jl}, for more up-to-date examples.

As described in the introduction, a QAOA is defined by three core components: a cost Hamiltonian, a mixer Hamiltonian, and a set of angles. 
For \juliq{}, the cost and mixer Hamiltonians are pre-computed and then passed to the simulator along with angles.
\juliq{} includes several common cost-functions, e.g. MaxCut, $k$-SAT, $k$-Densest Subgraph, etc. 
These all take as input some structure, usually a graph or a set of clauses, as well as a computational basis state (passed as an array of 0's and 1's) and output a scalar objective value.
Any user-defined cost function following this basic format will work as well.
See Listing~\ref{lst:basic} for a simple example of this approach. 

\begin{center}
\begin{minipage}{240pt}
\begin{lstlisting}[caption={Sample code for evaluating a  three round QAOA with MaxCut on an $n=6$ Erdos-Renyi graph.}, label={lst:basic}]
using JuliQAOA
using Graphs

# define graph
n = 6
graph = erdos_renyi(n, 0.5)

# calculate objective values across basis states
obj_vals = [maxcut(graph, x) for x in states(n)]

# generate mixer, [1] indicates \sum_i X_i
mixer = mixer_X([1], n)

p = 3 # number of rounds
angles = rand(2*p)
# angles[1:p] = betas, angles[p+1:2*p] = gammas

res = simulate(angles, mixer, obj_vals)

exp_value = get_exp_value(res)
\end{lstlisting}
\end{minipage}
\end{center}

The core \texttt{simulate()} command outputs a special object, which stores the statevector as well as objective values, and can be used to extract the expectation value, amplitudes for each state, and ground state probability.
The command \texttt{simulate\_gpu()} is equivalent in functionality to \texttt{simulate()} but runs on any available GPUs via \texttt{CUDA.jl} \cite{besard2018juliagpu}.

The unconstrained Pauli $X$ mixers are extremely efficient to calculate, and do not take much memory, so there is no need to save the results for future calculations.
The same cannot be said for constrained problems utilizing the Clique or Ring mixers.
In this case, \juliq{} provides an option to save the mixers to a user-specified location for re-use.
See Listing~\ref{lst:constrained} for an example of a constrained QAOA.
Note that in this case the cost function \texttt{densest\_subgraph} is called across all Dicke($n,k$) states, i.e. $n$-qubit states with Hamming weight $k$. 

\begin{center}
\begin{minipage}{240pt}
\begin{lstlisting}[caption={Setting up a Densest $k$-Subgraph problem with the Clique mixer. If the included file path exists, the pre-computed mixer is loaded. If it does not exist, the eigencomposition is stored for future re-use.},label=lst:constrained]
n = 6
graph = erdos_renyi(n, 0.5)

k = 3
obj_vals = [densest_subgraph(graph, x) for x in dicke_states(n, k)]

mixer = mixer_clique(n, k; file="path/to/saved/mixer")
\end{lstlisting}
\end{minipage}
\end{center}

A particular strength of \juliq{} is its flexibility and extensibility.
By default, \texttt{simulate()} begins the QAOA in the uniform superposition of all states (or all Hamming weight $k$ states, if the mixer is targeted at a specific weight).
However, \texttt{simulate()} accepts an optional argument \texttt{initial\_state}, which can be used to specified any other initial state, e.g. to study the effects of warm starts~\cite{egger2021warm}.
Furthermore, the \texttt{mixers} argument can accept an array of mixers (of length $p$) to test the efficacy of multiple mixers in a single QAOA.
In order to test multi-angle QAOA~\cite{herrman2021multiangle}, one can even pass an array of arrays of mixers, along with a nested array of angles, which allows for multiple mixers at each layer.

Finally, we include a simple call for finding good angles, see Listing~\ref{lst:optimize}.
\texttt{find\_angles()} defaults to attempting to maximize the objective value.
In the case of minimization, an overall minus sign can be added to the objective value list.
And in the case of problems where the objective values are both negative and positive, one must add an offset to make them all the same sign.

The function \texttt{find\_angles()} uses the angle-finding scheme described in~\cite{golden2021thresholdbased}, which starts by finding good angles at $p=1$ and then uses them as a seed for finding good angles at $p=2$, continuing up to a target number of rounds.
The results of each step of the angle-finding can be stored in a user-defined file.
If the angle-finding is interrupted for any reason, e.g. a server crash, it will load any saved results and resume from the last calculated angles.
At each step, we use the basinhopping algorithm, via \texttt{Basinhopping.jl}, to explore local minima.
The number of minima to explore, acceptance criteria, and other basinhopping parameters can be modified through optional arguments passed to \texttt{find\_angles()}.

\begin{center}
\begin{minipage}{240pt}
\begin{lstlisting}[caption={Examples of angle optimization outer-loop for $p$ rounds. \texttt{find\_angles()} defaults to an iterative approach, which uses the angles from round $p-1$ to seed the basinhopping at round $p$. If the included file path exists, the previously calculated angles will be used. One can also avoid the iterative approach by specifying \texttt{initial\_angles}, which will then perform basinhopping from that point. The function \texttt{find\_angles\_rand()} is included to showcase a straightforward user-defined angle-finding technique. In this case, we implement the random local minima search from~\cite{Lotshaw_2021}.}, label=lst:optimize]
find_angles(p, mixer, obj_values; file="path/to/saved/angles")

function find_angles_rand(p, mixer, obj_vals; iters=100)
    best_angles = []
    best_score = 0
    for _ in 1:iters
        x0 = 2*pi.*rand(2*p)
        res = optimize(x->-exp_value(x, mixer, obj_vals), x0, BFGS())
        if minimum(res) < best_score
            best_angles = minimizer(res)
            best_score = minimum(res)
        end
    end
    return best_angles
end

\end{lstlisting}
\end{minipage}
\end{center}

\begin{figure*}
  \centering
  \begin{subfigure}[b]{0.49\textwidth}
      \centering
      \includegraphics[width=\textwidth]{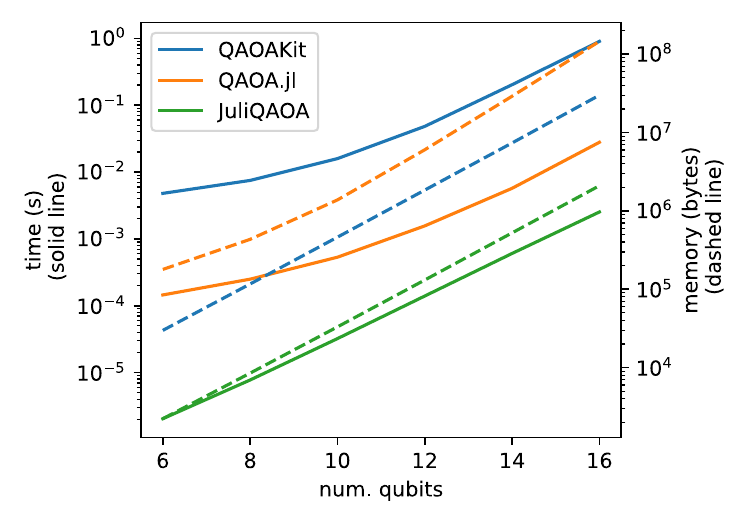}
      \caption{Scaling in qubits (with $p=1$)}
      \label{fig4_1}
  \end{subfigure}
  \hfill
  \begin{subfigure}[b]{0.49\textwidth}
      \centering
      \includegraphics[width=\textwidth]{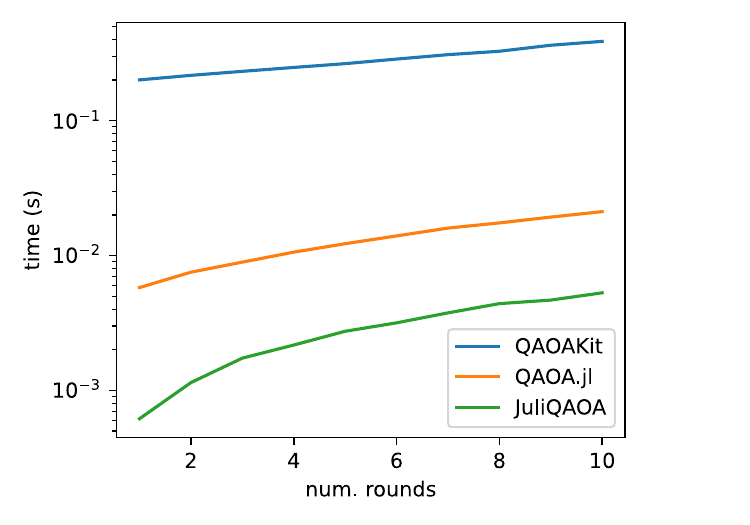}
      \caption{Scaling in rounds (with $n=14$)}
      \label{fig4_2}
  \end{subfigure}
     \caption{Performance comparison of \juliq{} against alternative QAOA simulation packages \qaoajl{} and \qaoakit{}. Figure~\ref{fig4_1} shows the scaling in qubit size for CPU time (solid lines) and memory usage (dashed lines) needed to simulate a $p=1$ MaxCut QAOA on a random $G(n,0.5)$ graph with the Transverse Field mixer. Figure~\ref{fig4_2} shows the scaling in rounds for CPU time needed to evaluate an $n=14$ MaxCut QAOA on a random $G(n,0.5)$ graph (memory increase is neglegible across rounds for all packages and is therefore not shown). All data generated on an Apple M2 Max device.}
     \label{fig2}
\end{figure*}

\section{Performance}
\label{section:Performance}
QAOA circuits can be written and simulated in many different simulation platforms, e.g. \texttt{Qiskit}~\cite{Qiskit} or \texttt{Pennylane}~\cite{bergholm2022pennylane}.
Two prominent QAOA-specific packages are the Python-based \qaoakit{}~\cite{Shaydulin_2021} and the recently introduced Julia package \qaoajl{}~\cite{Bode2023}.
\qaoakit{} only works with the MaxCut problem and transverse-field mixer, while \qaoajl{} can accommodate different mixers and problem types.
In both cases, these packages compose QAOA circuits and then pass them to general simulators for evaluation;
\qaoajl{} utilizes \texttt{Yao.jl}, while \qaoakit{} uses \texttt{Qiskit}.
We find that the pre-computation and purpose-built simulation of \juliq{} outperforms existing alternatives, see Figure~\ref{fig2}.
For example, for an $n=6, p=1$ MaxCut QAOA, \juliq{} is faster than \qaoakit{} by a factor of over 2000, and faster than \qaoajl{} by a factor of over 70.

\begin{figure}[!t]
    \centering
    \includegraphics[width=0.48\textwidth]{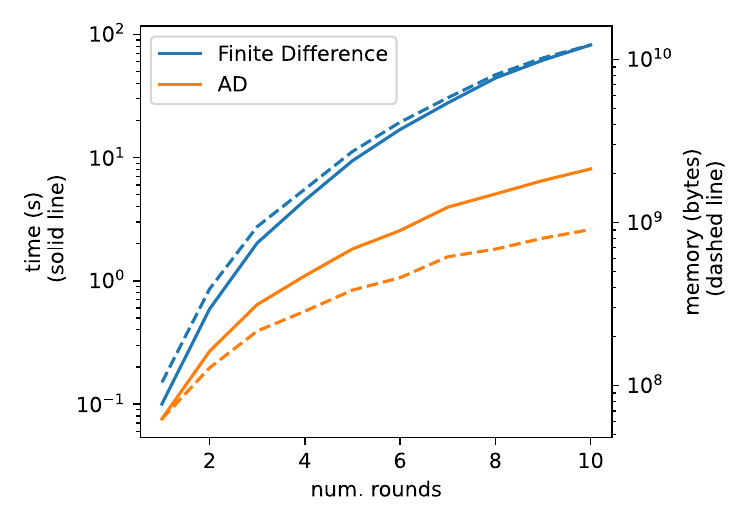}
    \caption{Performance comparison of local minima finding with the BFGS algorithm using either finite difference or AD to calculate gradients. Performance is averaged over 100 random $n=14$ MaxCut instances. All data generated on an Apple M2 Max device.}
    \label{fig3}
\end{figure}

Furthermore, the automatic differentiation built in to \juliq{} gives a factor of $O(p)$ improvement in time to compute the gradient of the QAOA expectation value $\braket{\boldsymbol{\beta}, \boldsymbol{\gamma}|C(x)|\boldsymbol{\beta}, \boldsymbol{\gamma}}$ in terms of $\boldsymbol{\beta}, \boldsymbol{\gamma}$.
This is because, after an initial caching pass, AD gives the gradient with a single evaluation of the expectation value and some constant overhead.
Meanwhile, traditional finite difference methods require at least $O(p)$ evaluations of the expectation value.
See Figure~\ref{fig3} for a comparison of time to find the closest local minima to a random initial point using the BFGS algorithm, with either finite difference or AD providing the gradient.

Perhaps most importantly, we find that \juliq{} facilitates a wide range of QAOA studies.
For example, only requiring a list of $C(x)$ evaluated across all feasible states allows total freedom in the choice of cost function, and simplifies usage for scientists without experience developing Hamiltonian encodings of optimization problems.
Researchers can explore arbitrarily complicated or synthetic optimization functions and mixer Hamiltonians, which can be useful in exploring the limits of QAOA performance.

Constrained optimization is a particular strength of \juliq{}.
In traditional circuit-based simulators, constrained optimization problems must be encoded in terms of unconstrained Hamiltonians with artificial ``penalty'' terms meant to dissuade the classical angle-finding optimization from selecting non-feasible solutions.
The design of \juliq{} allows us to use mixers which stay within the feasible subspace and simply ignore all non-feasible states, increasing accuracy and significantly reducing computational effort.

While this manuscript was under preparation, a similar QAOA simulation toolkit, \texttt{QOKit}, was introduced~\cite{lykov2023fast}.
\texttt{QOKit} shares many of the same fundamental ideas with \juliq{}, namely, precomputation and efficient implementation of mixer Hamiltonians.
However, there are several key differences between the two packages.
First, \texttt{QOKit} is written in Python, and leverages multiple backends for executing the statevector simulation (most notably NVIDIA's cuQuantum framework).
They have further optimized their code for large-scale GPU simulation, going up to $n=40$ with over 1000 GPUs. 
However, for exact statevector simulation they only support the Transverse Field mixer.
They include both Clique and Ring mixers, but their implementation is equivalent to a first-order Trotter approximation.
Finally, they do not currently provide support for automatic differentiation.

\begin{acks}
This work was supported by the U.S. Department of Energy through the Los Alamos National Laboratory. Los Alamos National Laboratory is operated by Triad National Security, LLC, for the National Nuclear Security Administration of U.S. Department of Energy (Contract No. 89233218CNA000001). Research presented in this article was supported by the NNSA’s Advanced Simulation and Computing Beyond Moore’s Law Program at Los Alamos National Laboratory. This material is based upon work supported by the U.S. Department of Energy, Office of Science, National Quantum Information Science Research Centers, Quantum Science Center. This research used resources provided by the Darwin testbed at Los Alamos National Laboratory (LANL) which is funded by the Computational Systems and Software Environments subprogram of LANL’s Advanced Simulation and Computing program (NNSA/DOE). This work has been assigned LANL technical report number LA-UR-23-28325.
\end{acks}

\bibliographystyle{ACM-Reference-Format}
\bibliography{references}

\end{document}